\documentclass[aps,prl,reprint,showpacs,longbibliography]{revtex4-1}

\usepackage{graphicx}
\usepackage{hyperref}
\usepackage{amssymb,amsmath}
\usepackage{color}
\usepackage{mathrsfs}

\newcommand{\eqr}[1]{Eq.~(\ref{#1})}

\newcommand{\figr}[1]{Fig.~\ref{#1}}

\newcommand{\braca}[1]{\left(#1\right)}
\newcommand{\bracb}[1]{\left[#1\right]}
\newcommand{\bracc}[1]{\left|#1\right|}

\newcommand{\bracga}[1]{\langle #1 \rangle}

\newcommand{\ket}[1]{|#1\rangle}

\newcommand{\var}{\mathrm{var}}

\newcommand{\bnu}{\boldsymbol\nu}

\newcommand{\bA}{\mathbf{A}}

\newcommand{\bD}{\mathbf{D}}
\newcommand{\bE}{\mathbf{E}}

\newcommand{\bM}{\mathbf{M}}
\newcommand{\bR}{\mathbf{R}}

\newcommand{\bW}{\mathbf{W}}

\newcommand{\bfm}{\mathbf{m}}

\newcommand{\nh}{\hat{n}}
\newcommand{\ph}{\hat{p}}

\newcommand{\xh}{\hat{x}}

\newcommand{\asc}{a_\mathrm{sc}}
\newcommand{\gid}{g_\mathrm{1D}}

\begin{document}

\title{Squeezing and entanglement of density oscillations in a Bose-Einstein condensate}
\author{Andrew~C.~J.~Wade}
\author{Jacob F.~Sherson}
\author{Klaus M\o{}lmer}

\affiliation{Department of Physics and Astronomy, Aarhus University, Ny
  Munkegade 120, DK-8000 Aarhus C, Denmark.}
\date{\today}

\begin{abstract}
The dispersive interaction of atoms and a far-detuned light field allows nondestructive imaging of the density oscillations in Bose-Einstein condensates. Starting from a ground state condensate, we investigate how the measurement back action leads to squeezing and entanglement of the quantized density oscillations. In particular, we show that properly timed, stroboscopic imaging and feedback can be used to selectively address specific eigenmodes and avoid excitation of non-targeted modes of the system.
\end{abstract}

\maketitle

Nonclassical, squeezed and entangled states of light provided by nonlinear optical components are important ingredients in quantum metrology and communication \cite{Wu1986a,Furusawa1998a}. Production of analogous states of matter have been a long quest in physics with implementations demonstrated in atomic ensembles, as well as in superconducting and nanomechanical devices. Relying on nonlinearities due to atomic interactions, ultracold atoms have been prepared in two-mode entangled states of split atomic clouds \cite{Esteve2008a,Berrada2013a} and internal state components \cite{Riedel2010a,Gross2011a,Hamley2012a,LŸcke2011a,Gross2010a,Muessel2014a}. Furthermore, it has been possible to prepare states that witness fundamental quantum properties through violation of inequalities obeyed by classical models
\cite{Vogels2002a,Perrin2007a,Keller2014a,Kheruntsyan2012a,LewisSwan2014a,Lopes2015a}.

Using dispersive light-matter interactions and measurement back action, room temperature vapour experiments have demonstrated squeezed \cite{Kuzmich2000a} and entangled \cite{Julsgaard2001a} states, quantum teleportation \cite{Sherson2006a}, and a quantum memory for light \cite{Julsgaard2004a}.
Similar experiments with interacting cold atoms and light fields have shown great progress \cite{Stamper1999a,Denschlag2000a,Saba2005a,Sanner2011a}, and numerous proposals exist 
\cite{Mekhov2012a,Hush2013a,Lee2014a} to manipulate the collective quantum states of ultracold atoms, while the full exploitation of the quantum mechanical nature of the interaction and measurements is yet to be realised.

\begin{figure*}[t]
  \centering
  \includegraphics[width=\textwidth]{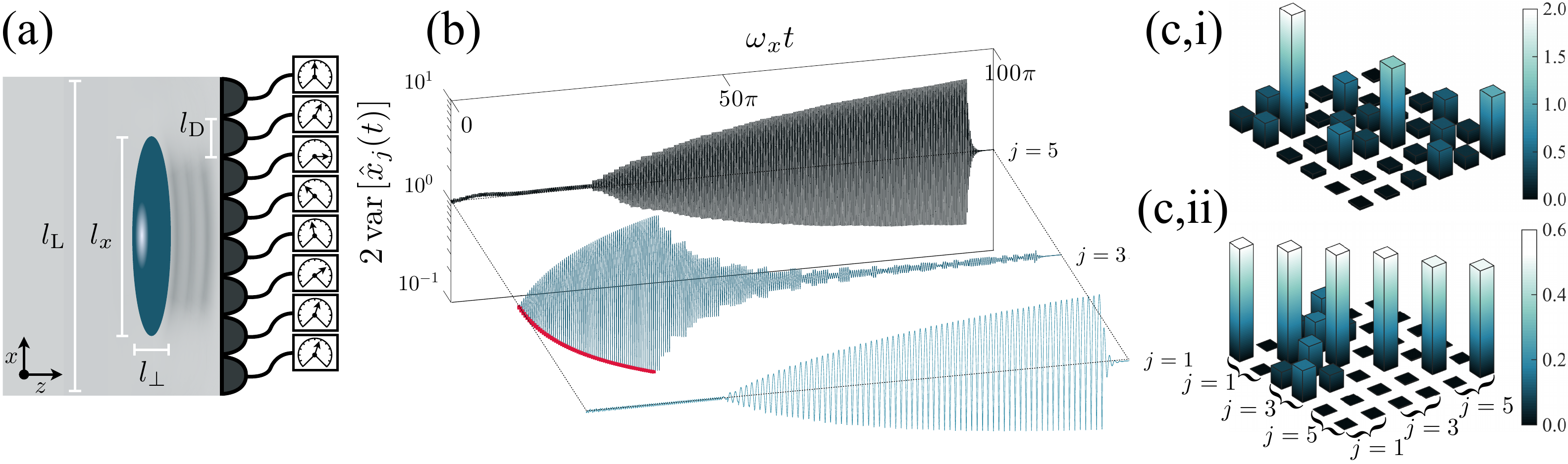}
\caption{(Colour online) (a) A BEC imprints a phase shift on a planar coherent light field depending on the atomic density along the BEC axis. The phase shifts are spatially detected by an encapsulating ($l_\mathrm{L} > l_x$) array of homodyne detectors (pixels) of widths $l_\mathrm{D} = l_x/10$. (b) The atomic modes can be selectively addressed and squeezed, illustrated here, where first the $3^\mathrm{rd}$ mode, and subsequently, modes $1$ and $5$ are squeezed. (c) The absolute value of the covariance matrix elements are shown for the first three odd modes. Continuous probing (i) squeezes and correlates a swathe of atomic modes, while
stroboscopic probing (ii) can generate correlations between only selected modes. The same level of entanglement of modes $1$ and $3$ is achieved in (i) and (ii), while the strong squeezing and the addressing of $5^\mathrm{th}$ in (i) is absent in (ii).
The simulations are performed for probing the $D_2$ line ($\sigma_+$ polarised light) of $1000$ $^{87}$Rb atoms in the $\ket{F,F_z}=\ket{2,2}$ state with $\omega_x=2 \pi \times150\mathrm{Hz}$, $\omega_\bot=100\omega_x$, and $\mu = 2\hbar \omega_x$.
\label{fig:system_cartoon}}
\end{figure*}

In this Letter we develop a theoretical treatment of Bose-Einstein condensate (BEC) dynamics due to the dispersive interaction with an optical probe field, subsequently measured, and feedback [\figr{fig:system_cartoon}(a)]. Specifically, we investigate the selective preparation of squeezed and entangled states of the multimodal excitations of the BEC.
Due to the temporal evolution, the local atomic density is not a quantum non-demolition (QND) variable \cite{Braginskii1978a,Thorne1978a,Braginskii1980a}, but by applying stroboscopic measurements at selected times, we can address effective QND variables of selected eigenmodes of the BEC dynamics. 

We consider a 1D ultracold Bose gas \cite{Petrov2000a} harmonically confined with axial (radial) trapping frequency $\omega_{x(\bot)}$ and the length scale $l_{x(\bot)}=\sqrt{\hbar/m\omega_{x(\bot)}}$, where radially tight confinement restricts the low energy excitations to axial motion. The BEC ground state wave function $f^{\phantom{2}}_0(x)$ (taken to be real) is given by the 1D Gross-Pitaevskii equation
\begin{equation}
[H_1+\gid n_0(x)] f^{\phantom{2}}_0(x) = \mu f^{\phantom{2}}_0(x),\nonumber
\end{equation}
where the BEC meanfield density is $n_0(x)=N f^2_0(x)$, and the chemical potential, $\mu$, enforces the BEC population to the total number of atoms, $N$. $H_{1}$ is the single atom Hamiltonian and the 1D interaction strength is $\gid=2\hbar^2 \asc/ m l_\bot^2$ with the \textit{s}-wave scattering length $\asc$.

The elementary excitations of the BEC are the collective center-of-mass, breathing, and higher order modes,
where each mode, $j$, constitutes a quantum harmonic oscillator degree of freedom with (dimensionless) quadrature observables $\xh_j(t)$ and $\ph_j(t)$. 
Their frequencies, $\omega_j$, and wavefunctions, $f_j^\pm(x)$, are found by solution of the coupled Bogoliubov-de Gennes equations,
\begin{equation}
\bracb{\begin{array}{cc} 0 & \mathcal{L}_+ \\ \mathcal{L}_- & 0 \end{array}}\bracb{\begin{array}{c} f_j^+(x)\\ f_j^-(x) \end{array}} = \hbar \omega_j \bracb{\begin{array}{c} f_j^+(x)\\ f_j^-(x) \end{array}}, \nonumber
\end{equation}
where $\mathcal{L}_\pm=H_1-\mu + (2\pm1) \gid n_0(x)$.
These eigenmodes provide an expansion of the probed atomic density,
\begin{equation}
\nh(x,t) = n_0(x) + 2 \sum_{j}\sqrt{n_0(x)} f_j^-(x)\xh_j(t) + O(N^0),\label{eq:atomic_den}
\end{equation}
embodying the dynamic quantum fluctuations about the BEC meanfield.
As we shall see, stroboscopically probing the density enables mode selective squeezing [red line in \figr{fig:system_cartoon}(b)] and entanglement [heralded by nonzero off-diagonal elements in \figr{fig:system_cartoon}(c,ii)], and reinitialization of the modes to oscillator ground states [end of \figr{fig:system_cartoon}(b)].

The dispersive light-matter interaction is characterized by the coupling constant $\kappa = -\sqrt{d_0 \eta}$  with the atomic depumping rate $\eta$, and optical depth on-resonance, $d_0$ \cite{Hammerer2010a,WadeGaus}. As we image the spatial density (\ref{eq:atomic_den}) by optical phase shift measurements [\figr{fig:system_cartoon}(a)], the light field detection in a certain pixel is sensitive to (a combination of) the atomic variables $\xh_j(t)$. 
This is described by the atomic-pixel mode overlap integrals,
\begin{equation}
\bar{f}_{jd} = \int_d \!\! dx \!\! \int \!\! dx' \mathcal{K}_{1} (x-x')f^{\phantom{2}}_{0}(x')f^-_{j}(x'), \nonumber
\end{equation}
where $\int_d dx$ denotes integration over the domain of the $d^\mathrm{th}$ pixel.
The convolution with the measurement kernel \cite{Dalvit2002a,Szigeti2009a}, $\mathcal{K}_\alpha \braca{x} = \int \!\! dk e^{-\frac{( \alpha  l_\mathrm{R} k)^4}{64\pi^2}  } e^{i k x}/2\pi$, accounts for the resolution limit along the BEC axis associated with the diffraction of light propagating over distances $\sim l_\bot$ through the BEC. By smearing out spatial features smaller than $l_\mathrm{R} = (l_\bot \lambda)^\frac{1}{2}$, where $\lambda$ is the light wavelength, it prevents the addressing of the higher modes with shorter spatial variations (the wavelength of mode $j$ scales as $\sim l_x/j$).
We will also require, for our analysis of the correlations between modes, the atomic mode overlap integrals
\begin{equation}
\bar{f}^2_{jk} = \int \!\! dx \!\! \int \!\! dx'  \mathcal{K}_{\sqrt[4]{2}} (x-x') f^{\phantom{2}}_{0}(x)  f^-_{j}(x) f^{\phantom{2}}_{0}(x')f^-_{k}(x'). \nonumber
\end{equation}

Since the dominant contributions of the light-matter interaction are second-order in light and atomic quadratures, and the light field is subject to quadrature measurements, the atomic quantum state can be described by a gaussian Wigner function. At all times, the state is fully characterised by $[\bR]_j = \bracga{\hat{\bnu}_j}$ and $[\bA]_{jk}=\mathrm{cov}(\hat{\bnu}_j,\hat{\bnu}_k)$, the first and second moments of the atomic mode variables, $\hat{\bnu}=\bracb{\xh_1(t),\ph_1(t),\xh_2(t),\ph_2(t),\ldots}^T$, respectively. 
Following the methodologies of \cite{Madsen2004a}, the quantum backaction of the light field measurements of the $\ph$ quadrature in each pixel [represented by expectation values and random Wiener increments, $dW_d(t)$] results in the stochastic evolution of the first moments \cite{WadeGaus}
\begin{equation}
d\bR=-\bD \bR dt + \bA\bfm\, d\bW. \label{eq:MTDOUE}
\end{equation}
The harmonic rotations in all $\{\xh_j(t),\ph_j(t)\}$ phase spaces is represented by the block diagonal matrix, $\bD$, with $2\times 2$ blocks $[\bD]_{j}  =  \left[\begin{array}{cc} 0 & -\omega_j \\ \omega_j & 0 \end{array} \right]$. 
Each pixel probes a linear combination of modes, represented by a rectangular matrix of $2\times 2$ blocks  $[\bfm]_{jd} =-\sqrt{\frac{ l_\mathrm{L}}{ l_\mathrm{D}}}\left[
                                                      \begin{array}{cc}
                                                        0 & 2\kappa\bar{f}_{jd} \\
                                                        0 & 0 \\
                                                      \end{array}
                                                    \right]
$, and owing to the correlations among the modes ($\bA$), the measurement results, $d\bW=[0,dW_1(t),0,dW_2(t),\ldots]^T$, affect the modes in a correlated manner.
 
The covariance matrix (second moments) evolve as \cite{WadeGaus}
\begin{equation}
\dot{\bA} =  \bE - \bD \bA - \bA \bD^T - \bA \bM \bA, \label{eq:MRDE}
\end{equation}
where $\bM = \bfm{}\bfm^T$ and the matrix $\bE$ with $2\times 2$ blocks, $[\bE]_{jk}  = \left[
                                                                       \begin{array}{cc}
                                                                         0 & 0 \\
                                                                         0 & \kappa^2  l_\mathrm{L}\bar{f}^2_{jk} \\
                                                                       \end{array}
                                                                     \right]$.
$\bA$ evolves in a manner independent of the measurement outcomes. While this implies we can deterministically assess the squeezing and entanglement properties of the system, we recall the random measurement results determine the mean displacements, about which, the squeezing and entanglement occurs. We shall now first consider the covariance matrix solutions  of \eqr{eq:MRDE}, focusing on the squeezing and entanglement generation, and later return to the stochastic evolution of the mean displacements described by \eqr{eq:MTDOUE}.

Probing with a constant optical field strength, $\kappa$, squeezes $\hat{x}_j(t)$ at a rate $\nu_j = \kappa^2  l_\mathrm{L}  \bar{f}^2_{jj}$, while the conjugate quadrature $\hat{p}_j(t)$ is anti-squeezed. Since $\hat{x}_j(t)$ and $\hat{p}_j(t)$ are coupled by rotation at rate $\omega_j$, the squeezing may be effectively suppressed. This is reflected by \eqr{eq:MRDE} yielding a steady state solution where the $\hat{x}_j(t)$ variance
$\var[\hat{x}_j(t)]_{\mbox{\tiny SS}} \simeq [(1+4\bar{\nu}_j^2)^{\frac{1}{2}}-1]^{\frac{1}{2}}/2\sqrt{2}\bar{\nu}_j \leq 1/2$,
and $\bar{\nu}_j = \nu_j/\omega_j$. Only if appreciable squeezing occurs during a fraction of a full phase space rotation, will there be squeezing in the long time limit.

For a given density, the rate of squeezing can only be increase by increasing the atomic depumping rate, $\eta$. Alternatively, a squeezed state of an oscillator with quadrature $\xh_j(t)=\xh_j^0 \cos \omega_j t+\ph_j^0 \sin \omega_j t$, where the $\hat{x}_j^0$ quadrature is squeezed, can still be achieved by applying a temporally modulated field \cite{Vasilakis2014a}. Here, we implement stroboscopic probing with a sequence of constant intensity pulses determined by one or more frequencies, $\varpi_i$. 
The timings, $t_l$, and durations, $\tau_l$, of the pulses are determined by ensuring, for each frequency, $\varpi_i t$ is within $\Delta \varphi/2$ from a multiple of $2\pi$. 
Probing at the single frequency, $\varpi=2\omega_j$, amounts to a train of $n$ pulses centred on times $t_l=[0,\pi/\omega_j,2\pi/\omega_j,\ldots,n\pi/\omega_j]$ with identical durations, $\tau_l=\tau= \Delta \varphi / \varpi$. Such a field addresses, and squeezes, the $\pm \hat{x}_j^0$ quadrature, while avoiding its anti-squeezing at intermediate times. Thus, enabling squeezing well beyond the continuous case for the same strength, $\kappa$. For small $\Delta \varphi$, the effective QND probing of $\hat{x}_j^0$ is expected, and hence, the variance $\var[\hat{x}^0_j]_{\mbox{\tiny QND}}=1/2(1+2\nu_j \tau_T)$ \cite{Duan2000a} for the accumulated probing time $\tau_T = n \tau$. This is observed (red line) in \figr{fig:system_cartoon}(b), where the first quarter of the trace squeezes the $3^\mathrm{rd}$ mode ($\kappa^2 = 100\omega_x/2\pi$, $\varpi=2\omega_3$ and $\Delta \varphi /2\pi=0.05$).

The atomic multimodal system is far from the single mode squeezing picture. However, if no other mode, nor coupled correlation, has a rational frequency ratio to $\omega_j$, the pulse train only addresses, and squeezes, the $j^\mathrm{th}$ mode. 
This modal selectivity is illustrated in \figr{fig:system_cartoon}(b), where the variances of modes $1$ and $5$ are essentially unaffected during the probing of the $3^\mathrm{rd}$ mode.
Here, the atomic interactions provide an irregular spectrum of the lowest frequency modes \cite{Petrov2000a}, allowing separate addressing without crosstalk.
In light of this, evaluating the performance of squeezing and entangling operations, we will later assess the selectivity within the multimodal $j=1,3,5$ subspace, as well as, this subspace's isolation from the rest of the system.

The simultaneous squeezing of two modes, $j$ and $k$, is achieved with the two frequencies, $\varpi_1=2\omega_j$ and $\varpi_2=2\omega_k$, as featured in \figr{fig:system_cartoon}(b) for modes $1$ and $5$ after $\omega_x t=25\pi$ ($\Delta \varphi /2\pi=0.1$).
The probing sequence is now out-of-phase with the $3^\mathrm{rd}$ mode, and hence, its prior squeezing is progressively lost returning to initial vacuum values.
The final $5\%$ of the trace demonstrates reinitialization of all modes to vacuum fluctuations by switching to weak continuous wave probing $(\kappa^2 = 50\omega_x/2\pi$).
The preservation of vacuum fluctuations of nontargeted modes, loss of squeezing of the $3^\mathrm{rd}$ mode, and reinitialization of all modes to vacuum fluctuations, featured in \figr{fig:system_cartoon}(b), are demonstrations of the interplay between atomic dynamics, measurement strength, and stroboscopic probing, enacting a quantum eraser for a subset of, or all, atomic modes.

\begin{figure}[ht]
  \centering
  \includegraphics[width=0.43\textwidth]{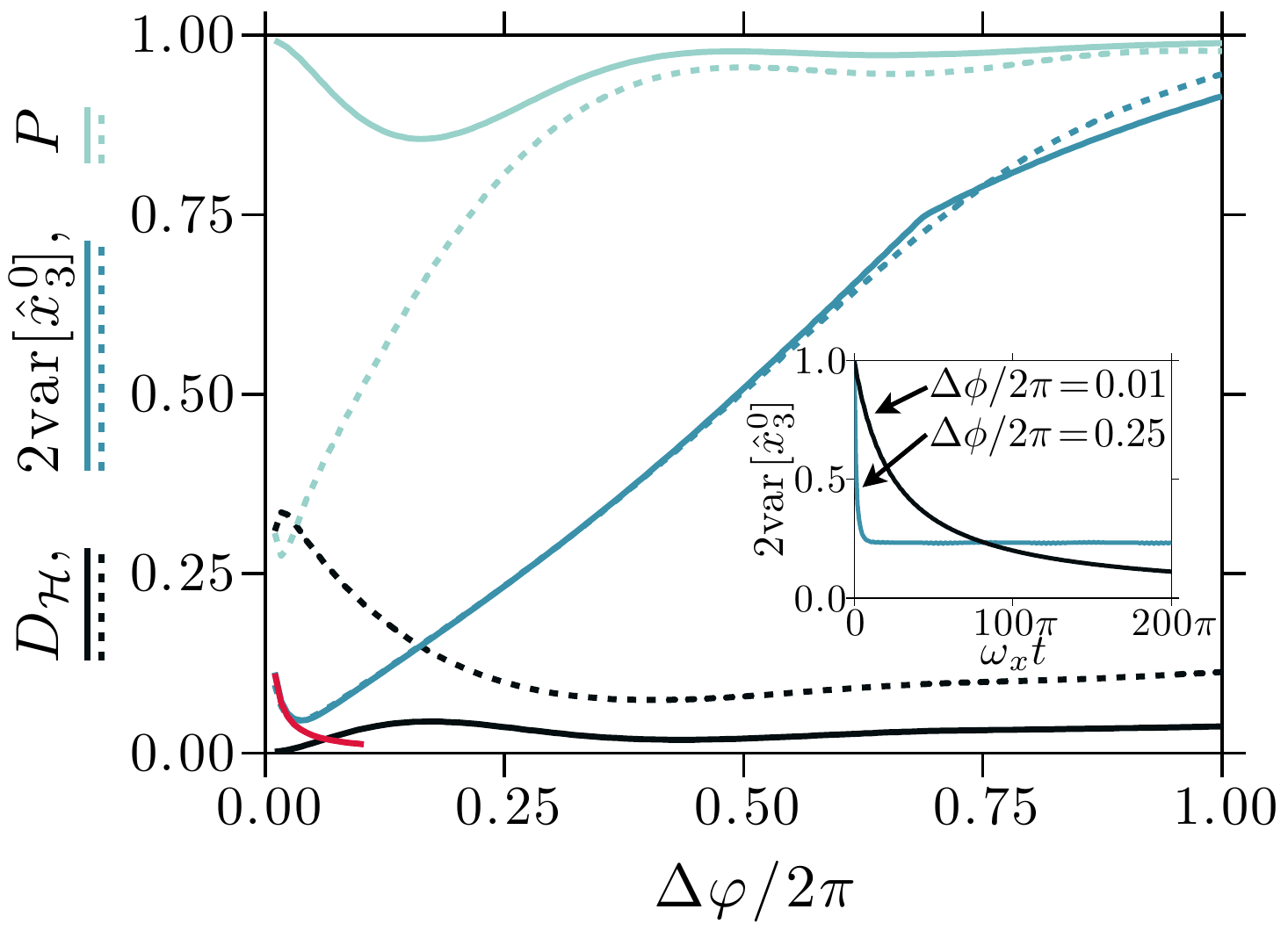}
  \caption{(Color online) The squeezing of the $3^\mathrm{rd}$ mode after stroboscopically probing ($\kappa^2 = 100\omega_x/2\pi$) for 100 trap periods is shown, while the temporal evolution of the quadrature variance, $\var[\hat{x}^0_3]$, is illustrated in the inset.
For small $\Delta \phi$ (pulse duration), $\var[\hat{x}_3^0]$ follows the QND result (red line, see text), while for larger $\Delta \phi$, the squeezing is faster but suboptimal. 
Excellent selectivity, Hellinger distance, $D_\mathcal{H}\simeq 0$, and little crosstalk, Purity, $P\simeq 1$, are observed for small $\Delta \phi$. The results for the noninteracting atomic system (dotted lines) are shown for comparison.
\label{fig:int_vs_nonint}}
   \end{figure}

To further assess the performance of the operations, we study squeezing of the $3^\mathrm{rd}$ mode in \figr{fig:int_vs_nonint} 
(similar results are found for other modes and for the joint squeezing or entangling of pairs of modes).
The rate of squeezing is determined by $\Delta \phi$, as the accumulated probing time $\tau_T \propto \Delta \phi$ (inset). 
However, only for smaller $\Delta \phi$, the $x$-quadrature is probed in a QND fashion (red line), while the squeezing for larger $\Delta \phi$ is suboptimal as a result of the inadvertent probing of the $p$-component.
The figure also addresses crosstalk between the $j=1,3,5$ subspace and its relative complement through the purity, $P$, of the reduced system density matrix, $\hat{\rho}$. The Hellinger distance \cite{marian2015a}, $D_\mathcal{H} =\mathrm{Tr}[\hat{\rho}^\frac{1}{2}-(\hat{\rho}_d)^\frac{1}{2}]^2 / 2$, quantifies the selectivity within the subset of modes. The desired state $\hat{\rho}_d$ assumes identical $\bR$ to $\hat{\rho}$, but with all blocks $[\bA]_{jk}$, except $j=k=3$, replaced by their initial vacuum values. Excellent selectivity ($D_\mathcal{H}\simeq 0$) and little crosstalk ($P\simeq 1$) are observed for small $\Delta \phi$. For comparison (dotted lines), a noninteracting atomic system demonstrates similar squeezing, however, the corresponding linear spectrum $\omega_j = \omega_x j$ results in significant crosstalk and poor mode selectivity.

\begin{figure}[ht]
  \centering
  \includegraphics[width=0.49\textwidth]{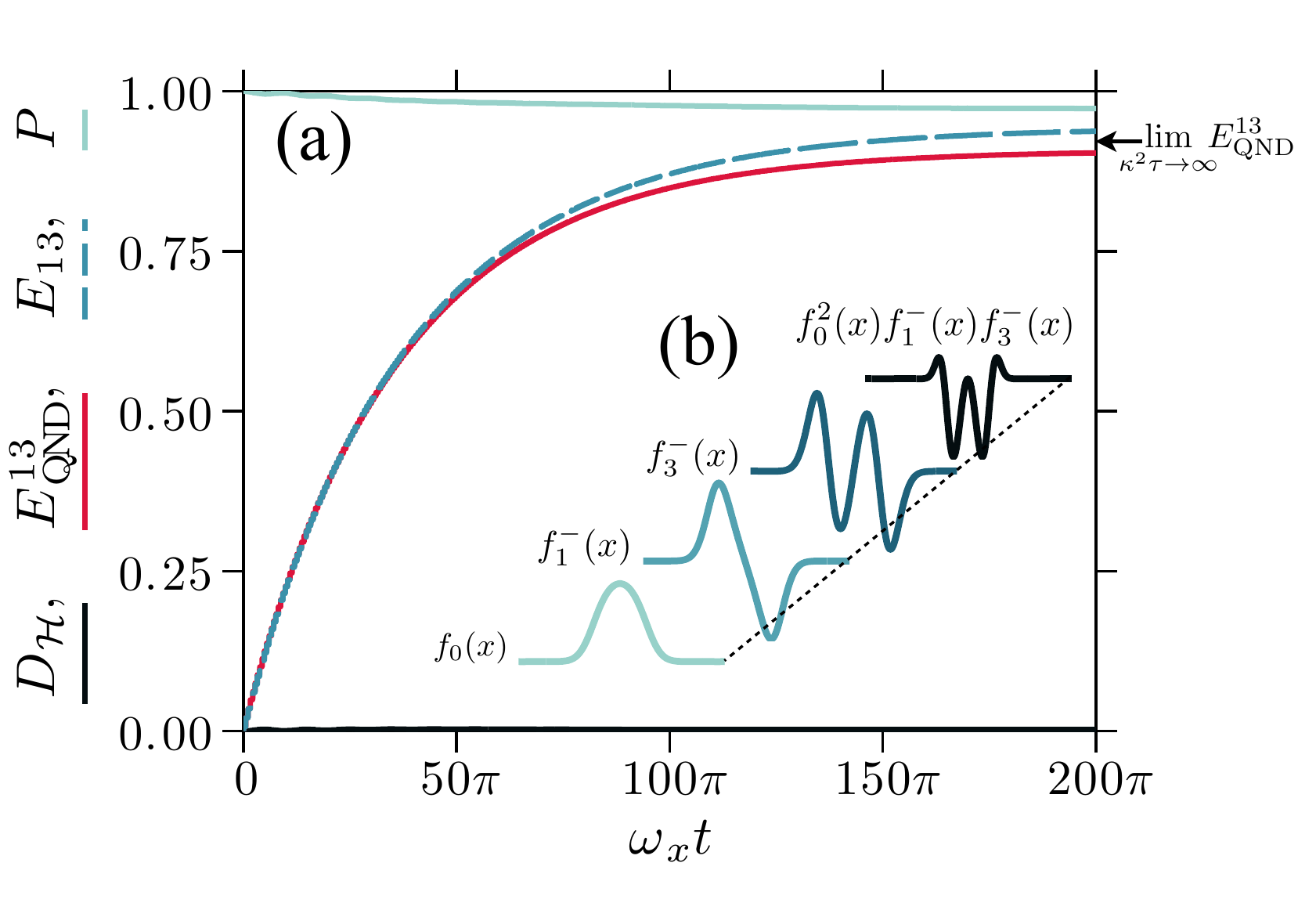}
  \caption{(Color online) (a) A pulse sequence of 100 trap periods with $\varpi_1=\omega_1 + \omega_3$, $\Delta \varphi/2\pi=0.03$, and $\kappa^2 = 30\omega_x/2\pi$, entangles modes $1$ and $3$.
 The logarithmic negativity $E_{13}$ slightly out-performs the corresponding QND result, $E^{13}_{\mathrm{QND}}$.
 The selectivity within the subspace is excellent ($D_\mathcal{H}\simeq 0$), while the subspace's purity $P\sim 97\%$. 
 (b) The mode functions (not to scale) of the condensate, $f^{}_{0}(x)$, entangled modes, $f^-_{1,3}(x)$, and the overlap, $\sim f^{2}_{0}(x)f^-_{1}(x)f^-_{3}(x)$.
\label{fig:ent_ex}}
\end{figure}

A probe transmitted through two atomic media reveals information about their collective rather than individual properties and may hence lead to their mutual entanglement \cite{Duan2000a}. Similarly, we may entangle two modes, $j$ and $k$, of a single BEC, by probing the density oscillation amplitude in a manner that does not discriminate contributions from the individual modes. The modes' respective spatial signatures, $\sim f^{\phantom{2}}_{0}(x)f^-_{j}(x)$ and $\sim f^{\phantom{2}}_{0}(x)f^-_{k}(x)$, must be indistinct ($\bar{f}^2_{jk}\neq0$). Partial temporal distinguishability, owing to different oscillation frequencies, $\omega_j$ and $\omega_k$, is avoided by stroboscopically probing with a train of pulses at the single frequency, $\varpi=\omega_j + \omega_k$. Analogous to the case of squeezing, it also allows selective addressing of modes as illustrated in \figr{fig:system_cartoon}(c). The absolute value of the covariance matrix elements of the $j=1,3,5$ subspace is shown for (i) the steady state of continuous probing ($\kappa^2 = 1000\omega_x/2\pi$) and (ii) stroboscopic probing for 100 trap periods ($\kappa^2 = 4\omega_x/2\pi$ and $\Delta \varphi /2\pi=0.03$). Although both cases reach the same level of entanglement of modes $1$ and $3$, the continuous probing addressing a swathe of modes, causes an additional squeezing, and requires a larger strength, $\kappa$.

Bipartite entanglement can be quantified by the logarithmic negativity $E_{jk}=\log_2\bracc{\bracc{\hat{\rho}^\mathrm{Tp}}}_\mathrm{Tr}$ \cite{Vidal2002a}, where the partial transpose and trace norm are denoted, $\mathrm{Tp}$, and  $\bracc{\bracc{\cdot}}_\mathrm{Tr}$, respectively. Figure \ref{fig:ent_ex}(a) shows the temporal build up of entanglement between modes $1$ and $3$, and the comparison to an effective QND probing that converges for $\kappa^2\tau \rightarrow \infty$ to the asymptotic limit, $E^{13}_{\mathrm{QND}} \rightarrow \log_4[\frac{1+\beta}{1-\beta}]$.
The spatial distinguishability of the mode functions, featured in \figr{fig:ent_ex}(b), remains the limiting factor, and is parameterised by $\beta = |\bar{f}^2_{jk}|/[\bar{f}^2_{jj}\bar{f}^2_{kk}]^{\frac{1}{2}}$.
Deviations from $E^{13}_{\mathrm{QND}}$ are attributed to coupling to other modes with commensurate correlation frequencies [here, $\omega_1 + \omega_3\simeq 2(\omega_1 + \omega_7)$] through $\bar{f}^2_{jk}$, as signalled by a small loss of purity ($P\sim 97\%$). The selectivity within the subspace is excellent ($D_\mathcal{H}\simeq 0$).

\begin{figure}[h]
  \centering
  \includegraphics[width=0.407\textwidth]{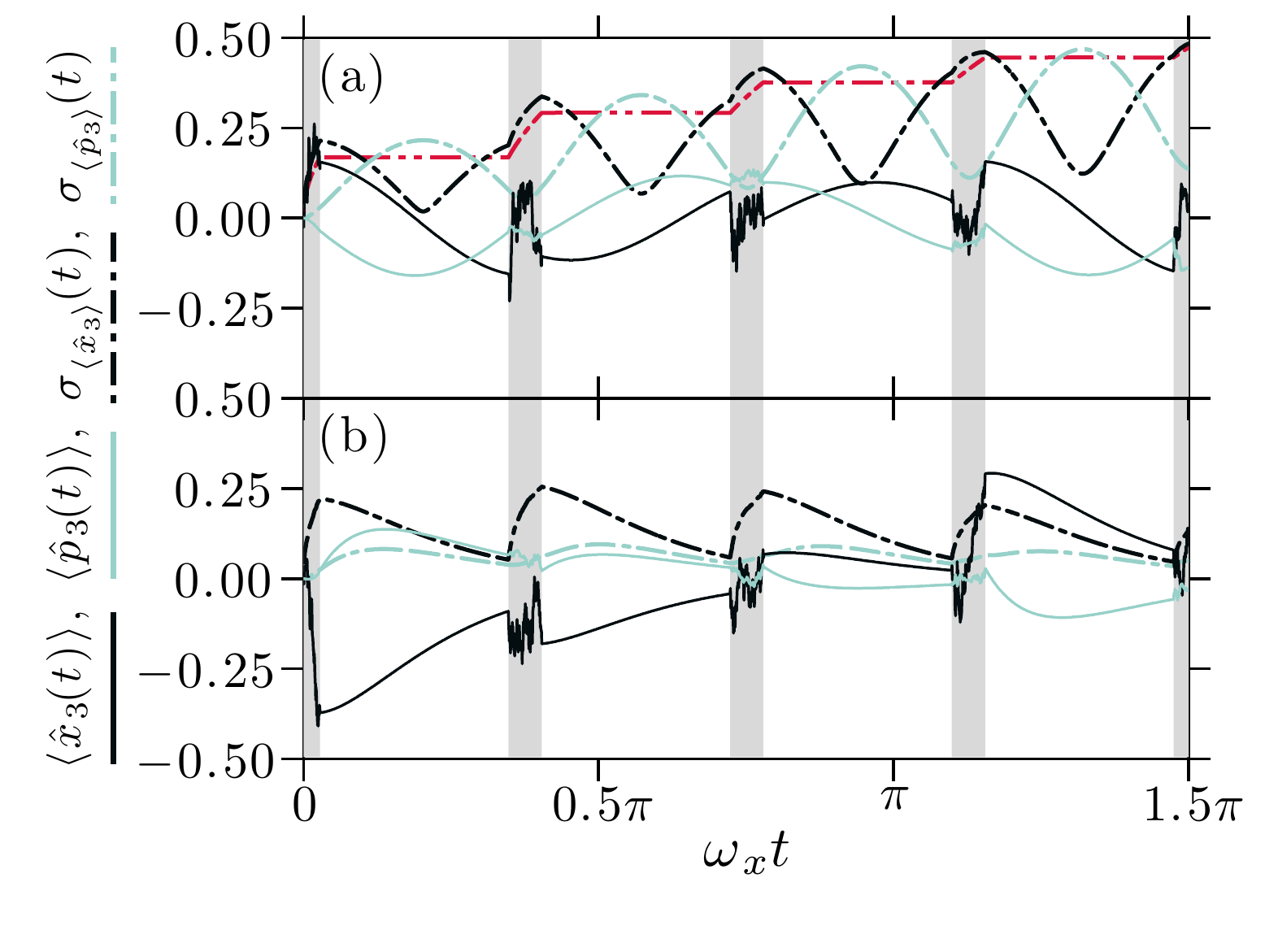}
  \caption{(Color online) A stochastically evolving trajectory of $\langle \hat{x}_3(t)\rangle$ and $\langle\hat{p}_3(t)\rangle$, and the standard deviations of $1000$ trajectories, $\sigma_{\langle \hat{x}_3\rangle}(t)$ and $\sigma_{\langle \hat{p}_3\rangle}(t)$, corresponding to the simulation in \figr{fig:int_vs_nonint} with $\Delta \phi /2\pi=0.15$. The grey regions indicate when the measurement occurs. Without feedback (a), the standard deviations of the trajectories can be approximated by $\sigma_{\langle \hat{x}_3\rangle}(t)=\sigma_{\langle \hat{p}_3\rangle}(t)=\sqrt{\nu_3 \tau_T /2}$ (red line). With feedback (b), we demonstrate the desired damping of mean values. \label{fig:traj_ex}}
\end{figure}

We have so far demonstrated that stroboscopic probing permits squeezing and entanglement of particular resonant modes. As illustrated in \figr{fig:traj_ex}(a), due to the probing backaction, the modes are also subject to random displacements, governed by the diffusion terms in \eqr{eq:MTDOUE}. The Gaussian multi-mode quantum state is, indeed, subject to such coherent displacement of all modes. This corresponds to a modification of the Gross-Pitaevskii mean field wave function, and it is suggestive that modulation of the trapping potential can be used as a feedback mechanism to recover and maintain the original shape $f_0(x)$. A micro-mirror array \cite{Goorden2014a} or spatial light modulator \cite{McGloin2003a} can provide such an adaptive lightshift potential for the atoms. Choosing the potential strength with single atom Hamiltonian, $H_\mathrm{f} = \hbar \sum_k h_k(x) \langle \hat{p}_k (t)\rangle$, where $h_k(x)=2\omega_j f_j^+(x) \delta_{jk} /\sqrt{n_0(x)} $, we can selectively address the displacement of the $j^\mathrm{th}$ mode. The corresponding action on the $j^\mathrm{th}$ mode is the addition of $\left[
                                                      \begin{array}{cc}
                                                        0 & 0 \\
                                                        0 & 2\omega_j \\
                                                      \end{array}
                                                    \right] $ to the $2\times 2$ block $[\textbf{D}]_{jj}$ in \eqr{eq:MTDOUE}. 
Due to the degenerate eigenvalue $-\omega_j$, the deterministic part of the evolution of $\hat{x}_j(t)$ and $\hat{p}_j(t)$ becomes critically damped and suppresses the displacements caused by the $d\textbf{W}$-terms in \eqr{eq:MTDOUE}. We have simulated the full multi-mode dynamics, and \figr{fig:traj_ex}(b) demonstrates the displacement of the squeezed mode in \figr{fig:int_vs_nonint} is successfully suppressed.

The feedback demonstrated in \figr{fig:traj_ex} is only performed for a single mode (although not limited to).
In the effective QND examples of single mode squeezing (e.g., $\Delta \phi=0.01$ in \figr{fig:int_vs_nonint}) and bipartite entanglement (\figr{fig:ent_ex}), the coherent excitation of mean values corresponds to an $\sim 8\%$ population $N_\mathrm{nc}$ outside of the BEC mode.
If necessary, more elaborate $H_\mathrm{f}$ can be investigated to minimise the excitation of multiple modes, and $N_\mathrm{nc}$ may be reduced by different trapping geometries, where the measurement kernel $\mathcal{K}_\alpha \braca{x}$ may yield less coherent excitation of the irrelevant modes. 

In conclusion, we have demonstrated the quantum control of a matter-wave system via spatially resolved optical probing. Using stroboscopic probing and feedback, we can address and correlate effective QND observables of selected density modes of a BEC, while preserving the initial vacuum fluctuations of nontargeted modes. The performance of our proposal is ultimately limited by residual atomic depumping \cite{Hammerer2004a} and by the depletion of the BEC ground state when many modes become significantly squeezed.
With current experimental parameters \cite{Widera2008a,Langen2013a,Haller2009a}, however, a small alkali BEC should undergo sufficient squeezing and entanglement to allow demonstration of a few-mode memory for scalable quantum information processing.

\begin{acknowledgments}
A.C.J.W.~and K.M.~acknowledge support from the Villum Foundation Center of Excellence, QUSCOPE, and the E.U.~Marie Curie program ITN-Coherence 265031. 
J.F.S.~acknowledges support from Lundbeckfonden and the Danish Council for Independent Research.
A.C.J.W.~wishes to thank C.K.~Andersen and M.C.~Tichy for fruitful discussions.
\end{acknowledgments}


\begin{thebibliography}{46}
\bibitem [{Wad()}]{WadeGaus}%
  \BibitemOpen
  \href@noop {} {}\bibinfo {howpublished} {Details to be presented
  elsewhere.}\BibitemShut {Stop}%
\end{thebibliography}
\end{document}